\documentclass[a4paper,final,12pt]{article}
\usepackage{graphicx}
\usepackage{amssymb,amsthm,amsmath}

\let\abs=\envert

\newcommand{\ket}[1]{\left\lvert#1\right\rangle}

\newcommand{\scalop}[3]{\langle #1 \vert #2 \vert #3 \rangle}

\newcommand{\lagr}{{\cal L}}

\newcommand{\zm}[1]{\overset{o}{#1}}
\newcommand{\Zm}[1]{\left \langle #1 \right \rangle_0}
\newcommand{\nm}[1]{\overset{n}{#1}}
\newcommand{\Nm}[1]{\left \langle #1 \right \rangle_n}

\newcommand{\A}{{\mathbf A}}
\newcommand{\F}{{\mathbf F}}
\newcommand{\der}{{\mathbf D}}
\newcommand{\V}{{\mathbf V}}
\newcommand{\J}{{\mathbf J}}

\newcommand{\PHI}{{\boldsymbol\Phi}}

\newcommand{\tr}{{\text{Tr}}}
\newcommand{\vp}{\varphi}
\newcommand{\la}{\lambda}

\newcommand{\Z}{\hbox{{\rm Z\hskip-4ptZ\hskip1pt}\,}} 
\newcommand{\HZ}{{\rm I\!H}}

\newenvironment{rmenum}{
	
	\begin{enumerate}
     }{
	\end{enumerate}
	
     }

\title{A model for $SU(3)$ vacuum degeneracy using light-cone coordinates}

\author{Gr\'{e}gory Soyez\\
Institut de Physique, B\^{a}t B5\\
Universit\'{e} de Li\`{e}ge, B-4000 Li\`{e}ge, Belgium\\
{\em G.Soyez@ulg.ac.be}}

\begin{document}

\maketitle

\begin{abstract}
\noindent Working in light-cone coordinates, we study the zero-modes and the vacuum
in a $2+1$ dimensional $SU(3)$ gauge model. Considering the fields as independent of the tranverse variables, we dimensionally reduce this model to $1+1$ dimensions.
After introducing an appropriate $su(3)$ basis and gauge conditions,
we extract an adjoint field from the model. Quantization of this adjoint field and
field equations lead to two constrained and two dynamical zero-modes.
We link the dynamical zero-modes to the vacuum by writing down a Schr\"{o}dinger
equation and prove the non-degeneracy of the $SU(3)$ vacuum provided that we neglect the contribution of constrained zero-modes.
\end{abstract}

\section*{Introduction}

\hspace{0.5cm} In this paper, we quantize the pure-gauge sector of QCD in $2+1$ dimensions. It is hoped that this is a step towards the solution of the real world case, which at present seems too complicated to tackle. More precisely, we study the vacuum degeneracy. Working with a degenerate vacuum, as it is the case when studying QCD in front form, has always been a problem in quantum field theory. Actually, the property of vacuum triviality in light-cone coordinates should give a solution to that problem. In axiomatic field theory, it can be shown that the existence of a probability density and of a complete set of states implies vacuum triviality. With that point of view, proving unicity of the QCD vacuum should constitute another argument for QCD as a coherent quantum field theory.

Some work has already been done in this direction. One of the most interesting is the study of the vacuum degeneracy in a $2+1$ dimensional model for $SU(2)$. This model, developed by A. Kalloniatis, can be found in the two papers \cite{Kallo1, Kallo2}. The first steps towards a generalization including fermionic fields have been performed in \cite{tachi}. The present paper is a generalization to the $SU(3)$ case of Kalloniatis's model. In fact, it continues in a natural way the evolution of the study of the QCD vacuum degeneracy in light-cone coordinates.
All calculations will be performed with fields defined in $2+1$ dimensions but our model will be dimensionally reduced to $1+1$ dimensions. Going from $2+1$ dimensions to $3+1$ replaces a second order differential equation with a system of two coupled equations, complicating all calculations. 

We use light-cone coordinates to perform this study. These coordinates, introduced by
Dirac \cite{dirac} in 1949, are usual in high energy physics, due to their natural matching with the light-cone. 
Moreover, it is well known that light-cone coordinates
are adapted to the study of the vacuum. In order to
work in a Hamiltonian formalism and to avoid boundary condition problems, we adopt
the point of view of Discretized Light-Cone Quantization (DLCQ) \cite{dlcq1, dlcq2, bible}.
This means that we impose a periodicity condition $\phi(x^-=-L) = \phi(x^-=L)$ for
every field $\phi$.

We will face the problem of gauge fixing. In abelian gauge theory, gauge fixing is
complicated by the problem of constrained zero-modes \cite{abel1, abel2, abel3, abel4}.
When studying non abelian gauge theories, we also have to deal with the extra complication of dynamical zero-modes
\cite{nonab, nonab2}. These are coming from the fact that we are working with the non trivial
topology of a hyper-torus. Roughly speaking, we have to replace the light-cone gauge $\A^+=0$ by
$\partial_- \A^+ = 0$ and additional conditions in the zero-mode sector. With our choice of gauge conditions, we will have to deal with $2$ constrained zero-modes and $2$ dynamical ones. While constrained zero-modes \cite{contr} preserve the vacuum triviality, as for example in $\phi^4$ theory \cite{phi4-1, phi4-2, phi4-3, phi4-4}, dynamical zero-modes are susceptible to give rise to vacuum degeneracy.

In order to simplify calculations, we will make the assumption that all fields are independent of transverse variables. This allows us to perform a dimensional reduction. Moreover, we won't solve the constraints on the 2 constrained zero-modes in this paper. Hence, we simply neglect their contribution to the vacuum wave-function.

The constraints being linear, this model should not lead to any spontaneous breaking of $SU(3)$ symmetry. Writing down a Shrödinger equation for the light-cone vacuum, we will finally conclude to its triviality.

In this paper, we follow the conventions introduced by 
Kogut and Soper \cite{KS} by introducing $x^{\pm} = \frac{1}{\sqrt{2}} (x^0 \pm x^1)$. We consider
$x^+$ as our evolution para\-me\-ter, while $x^-$ is a longitudinal variable. 

This paper is structured as follow : in section 1, we will write down field equations in a general $SU(N)$ gauge theory and introduce the scalar adjoint fields. Section 2 introduces a suitable choice of $su(3)$ basis. The notion of zero-modes and normal modes in DLCQ is introduced in section 3. Section 4 sets our gauge conditions and rewrites the fields equations. DLCQ being easier in a second quantization formalism, section 5 establishes Fock expansions for the adjoint field. The existence of Gribov copies, leading to the choice of a fundamental domain, is discussed in section 6. Section 7 is devoted to the resolution of Gauss's law which allows us to impose conditions on the physical states and to associate quantum numbers with the adjoint field. Section 8 and 9 study respectively the impact of constrained and of dynamical zero-modes on the theory. We conclude in section 10.

\section{$SU(3)$ Gauge theory and Lagrange equations}

\hspace{0.5cm} We consider a gauge vector field $\A^\mu$ with values in $su(N)$ and defined in $d+1$ dimensions. The covariant derivative is then given by
\begin{equation}
\der^\mu = \partial^\mu + ig [\A^\mu, \cdot].
\end{equation}
As usual, we can define a chromomagnetic tensor $\F^{\mu\nu}$.

The usual $SU(N)$ Yang-Mills Lagrangian can be written
\begin{equation}\label{TheL}
{\cal{L}} = -\frac{1}{2} \tr(\F^{\mu\nu}\F_{\mu\nu}) = -\frac{1}{4} F^{\mu\nu}_a F^a_{\mu\nu},
\end{equation}
with $\F^{\mu\nu} = F^{\mu\nu}_a \tau^a$ and $F^{\mu\nu}_a = \partial^\mu A^\nu_a - \partial^\nu A^\mu_a - g {f_a}^{bc} A^\mu_b A^\nu_c$.

For convenience, in light-cone coordinates, the Greek indices $\alpha$, $\beta$, ... are running over $+$ and $-$, while Latin indices take the values $2, 3, \dots, d$.
Therefore, the stress-energy tensor can be computed. It reads
\begin{equation}\label{Stress}
T^{\mu\nu} = 2 \tr(\F^{\mu\kappa}\F_\kappa^{\;\;\nu}) - g^{\mu\nu} \lagr,
\end{equation}

Now and in the following, we restrict our model by assuming that fields are independent of transverse variables, as done by Kalloniatis in \cite{Kallo1} for the $SU(2)$ case. Mathematically, this means
\begin{equation}\label{trg}
\partial_i \A^\mu = 0, \quad \quad \forall \mu= +,-,2,\dots,d,\quad \forall i=2,\dots,d.
\end{equation}
In other words, the field $\A^\mu$ is assumed to depend only on $x^+$ and $x^-$.

Following a regularization in supersymmetric theories \cite{Siegel1, Siegel2}, we introduce the notation
\[
\A^\mu \equiv (\A^+, \A^-, \A^i) = (\V, \A, \PHI^i),
\]
where the fields $\PHI^i$ are called {\em scalar adjoint fields}. This step is sometimes called ``dimensional reduction'' because we are left with a $1+1$ dimensional model with $d-1$ adjoint fields.

Before writing down fields equations, let us restrict the dimension to $2+1$ dimensions, reduced to $1+1$, in such a way that only one adjoint field $\PHI$ is needed. The Lagrangian thus becomes
\begin{equation}
\lagr = - \frac{1}{2} \tr(\F^{\alpha\beta} \F_{\alpha\beta}) + \tr(\der^\alpha \PHI \der_\alpha \PHI).
\end{equation}
A straightforward calculation shows that Euler-Lagrange equations can be put in the form
\begin{eqnarray}
\der_\alpha \F^{\alpha\beta} & = & g \J_M^\beta, \label{mvtAV}\\
\der^\alpha\der_\alpha \PHI  & = & 0, \label{mvtPHI}
\end{eqnarray}
where $\J_M^\alpha = -i [\PHI, \der^\alpha\PHI]$ is called the {\em matter current}. The field $\PHI$ can be seen as the source of the fields $\V$ and $A$. Eq. \eqref{mvtAV} shows that this coupling occurs through the matter current.

Although these equations hold in $SU(N)$, we only consider them in the special case of $SU(3)$ which is the QCD gauge group.

\section{$SU(3)$ conventions}

\hspace{0.5cm} Having restricted the gauge group to $SU(3)$, we now fix our conventions about the $su(3)$ algebra. In order to simplify the quantization, we won't use the Gell-Mann representation. The exact representation for the generators is given in appendix A. That choice of matrices is a new way to approach a $SU(3)$ gauge model.
The matrices constituting this representation can be obtained from Gell-Mann matrices by transformations very similar to the ones defining light-cone coordinates. For example, we have
\begin{eqnarray*}
\lambda^1 = \frac{1}{2\sqrt{2}}(\lambda^1_{GM} + i \lambda^2_{GM}),
\lambda^2 = \frac{1}{2\sqrt{2}}(\lambda^1_{GM} - i \lambda^2_{GM}),
\end{eqnarray*}
where the $GM$ subscripts refer to Gell-Mann matrices. We have the same transformations for $(\lambda^4, \lambda^5)$ and for $(\lambda^7, \lambda^8)$. See the appendix for details and reasons motiving this particular choice of basis.

\section{Zero-modes and normal-modes}

\hspace{0.5cm} In order to solve the field equations and to study their influence over the vacuum structure, we adopt the point of view of DLCQ. This means that we impose the periodicity condition $\phi(x^-=-L) = \phi(x^-=L)$ for every field $\phi$. The {\bf zero-mode} of such a field is defined by
\begin{equation}
\Zm{\phi} = \zm{\phi} \equiv \frac{1}{2L} \int_{-L}^L \phi(x) \, dx^-.
\end{equation}
In the same way, the {\bf normal modes} of $\phi$ are
\begin{equation}
\Nm{\phi} = \nm{\phi} \equiv \phi - \zm{\phi},
\end{equation}
where $n$ stands for {\em normal}.

Physically, the zero-mode $\zm{\phi}$ can be interpreted as the Fourier component of $\phi$ with vanishing momentum $P^+$.

\section{Gauge conditions}

\hspace{0.5cm} We still can use gauge freedom to simplify equations \eqref{mvtAV} and \eqref{mvtPHI}. Following the $SU(2)$ case \cite{Kallo1}, we choose
\begin{enumerate}
\item $\partial_- \V = 0$ or $\nm{\V} = 0$. This is the usual light-cone gauge\footnote{This gauge condition is used for example in lattice calculations and to describe light-cone wave-functions.}, given that the condition $\V=\A^+=0$ cannot be satisfied in general. This gauge condition reduces $\V$ to a zero-mode, depending only on $x^+$.

\item By performing a $SU(3)$ transformation, $\V$ can be diagonalized and written
\begin{equation}
\V(x^+) = v_3(x^+) \lambda^3 + v_8(x^+) \lambda^8.
\end{equation}
It is very useful for following developments to introduce $z_3 = \frac{gLv_3}{\pi}$ and $z_8 = \frac{gLv_8}{\pi}$.

\item As in QED or in $SU(2)$, replacing the condition $\A^+=0$ by $\partial_- \A^+=0$ leaves gauge freedom in the zero-mode sector. In this sector, we shall set $\zm{A_3} = 0$ and $\zm{A_8}=0$.

\item We must stress the fact that there is still a residual gauge freedom : the one generated by all transformations conserving the diagonal form of $\V$. We will see later that these redundancies can be eliminated by imposing conditions on physical states.
\end{enumerate}

Let us see what kind of simplifications on fields equations are implied by these conditions. First of all, equation \eqref{mvtAV} becomes for $\beta = +$,
\begin{equation}\label{eqnA}
-\der^2_- \A = -g \J^+_M,
\end{equation}
On the other hand, for $\beta=-$, we have 
\begin{equation}
\partial_+ \partial_- \A - \partial_+^2 \V + ig [\A, \partial_- \A] 
              -ig[\A, \partial_+ \V] - g^2 \left[\A,[\V,\A] \right] = g \J^-_M. \label{ev}
\end{equation}
This last equation gives the time evolution for the fields $v_3$ and $v_8$ (or, equivalently, $z_3$ and $z_8$). These fields are thus dynamical zero-modes. Rather than solving this equation, we will consider $v_3$ and $v_8$ as simple variables. 
We will come back later to equation \eqref{mvtPHI}.

Next, we can consider the stress-energy tensor. We deduce the Hamiltonian and the longitudinal impulsion operator from \eqref{Stress}
\begin{eqnarray}
P^- & = & \int_{-L}^L dx^- \, T^{+-} 
      =   \int_{-L}^L dx^- \, \tr\left((\partial_+\V - \der_- \A)^2\right), \nonumber \\
P^+ & = & \int_{-L}^L dx^- \, T^{++} 
      =   \int_{-L}^L dx^- \, 2 \tr \left( (\der_-\PHI)^2\right).\nonumber
\end{eqnarray}
We also introduce the following dimensionless quantities
\begin{equation}\label{hamilt}
\hat{K} = \frac{4\pi^2}{g^2L} P^+ \quad \quad \text{et} 
                                  \quad \quad \hat{H} = \frac{4\pi^2}{g^2L} P^-.
\end{equation}

\section{Quantization}

\hspace{0.5cm} In this section, we shall quantize the adjoint field, constrained by equation \eqref{mvtPHI}.
The field conjugate to $\PHI$ is found to be
\begin{equation}\label{pidef3}
{\boldsymbol \pi} = \partial_- \PHI - ig [\PHI, v_3\la^3+v_8\la^8].
\end{equation}
Using our $su(3)$ basis, $\PHI = \vp_a \la^a$, and hermiticity, we get
$\vp_2 = \vp_1^\dagger$, $\vp_5 = \vp_4^\dagger$ and $\vp_7 = \vp_6^\dagger$.
Equation \eqref{pidef3} explicitly becomes 
\begin{eqnarray}\label{phicomp}
&&\begin{cases}
\pi_3 = \partial_- \vp_3, \\
\pi_8 = \partial_- \vp_8,
\end{cases} \nonumber \\
&&\begin{cases}
\pi_1 = (\partial_-+igv_3) \vp_1,\\
\pi_2 = (\partial_--igv_3) \vp_2, 
\end{cases} \nonumber \\
&&\begin{cases}
\pi_4 = (\partial_-+\frac{ig}{2}v_3+igv_8) \vp_4,\\
\pi_5 = (\partial_--\frac{ig}{2}v_3-igv_8) \vp_5,
\end{cases} \nonumber \\
&&\begin{cases}
\pi_6 = (\partial_--\frac{ig}{2}v_3+igv_8) \vp_6,\\
\pi_7 = (\partial_-+\frac{ig}{2}v_3-igv_8) \vp_7.
\end{cases}
\end{eqnarray}

Quantization of $\vp_3$ and $\vp_8$ is at all levels similar to quantization of a single scalar field. Therefore, the Fock expansion for these fields is
\begin{equation}
\vp_j = \frac{a_{0,j}}{\sqrt{4\pi}} 
      + \sum_{n=1}^\infty \frac{w_n}{\sqrt{4\pi}} \left(
	      a_{n,j} e^{-i\frac{n\pi}{L}x^-}
	    + a_{n,j}^\dagger e^{i\frac{n\pi}{L}x^-} \right)
\quad \quad \quad \quad j=3,8
\end{equation}
with $w_n = \frac{1}{\sqrt{n}}$. For $i,j = 3.8$ and $k,l \geq 1$, the only non-vanishing commutators are 
\begin{equation}
\lbrack a_{k,i}, a_{l,j}^\dagger \rbrack = \delta_{kl} \delta_{ij}.
\end{equation}

The case of off-diagonal elements of $\PHI$ is much more complicated \cite{quant1,quant2}. Actually, it is easy to see from equation \eqref{phicomp} that we can group the six off-diagonal components of $\PHI$ into three pairs. Each of these three pairs of fields may be quantized separately and directly deduced from the $SU(2)$ case treated in \cite{Kallo1}. We summarize here the steps leading to the quantization of $\vp_1$ and $\vp_2$. Given the periodicity condition, our starting point will of course be a Fourier expansion
\begin{eqnarray}
\vp_1 & = & \sum_n C_n^\dagger e^{i\frac{n \pi}{L} x^-}, \label{Four1}\\
\vp_2 & = & \sum_n C_n e^{-i\frac{n \pi}{L} x^-}
\end{eqnarray}
where the summation runs over $n \in \Z$ and the fact that $\vp_2 = \vp_1^\dagger$ is used.
Calculating conjugate fields through equation \eqref{phicomp} and inserting the result into the canonical commutation relation\footnote{Note that, in general, $\pi_1$ and $\pi_2$ will have a zero-mode due to the second term in the covariant derivative. Thus, we may use the full canonical commutation relation and not the one restricted to normal modes as it is the case while quantizing $\vp_3$ or $\vp_8$.}
\begin{equation}
[\vp_-(x), \pi^-(y)] = \frac{i}{2} \delta(x^--y^-),
\end{equation}
we obtain
\begin{equation}
[C_m, C_n^\dagger] = \frac{1}{4\pi (n+z_3)} \delta_{nm} 
                   = \frac{\text{sign}(n+z_3)}{4\pi \abs{n+z_3}} \delta_{nm}.
\end{equation}


All the work reside then in transforming the field development in order to bring back usual commutation relation between creation and destruction operators. The trick is to transform the sum over all integers into a sum over half-integers. Thus, we introduce $\HZ = \Z+\frac{1}{2} = \{\pm \frac{1}{2}, \pm \frac{3}{2}, \dots \}$ and rewrite \eqref{Four1} as
\footnote{Introduce $m=n+m_0$ and $\tilde{A}_m = C_{m-m_0} \sqrt{4\pi\abs{m-m_0+z_3}}$.}
\begin{equation}\label{vpm2}
\vp_2 = \frac{e^{i\frac{m_0 \pi}{L} x^-}}{\sqrt{4\pi}} 
        \sum_{m\in \HZ} \frac{\tilde{A}_n}{\sqrt{\abs{m+z_3-m_0}}} e^{-i\frac{m \pi}{L} x^-}.
\end{equation}
This relation is valid for every $m_0 \in \HZ$ and leads to the commutation relation $[\tilde{A}_m, \tilde{A}_n^\dagger]=\delta_{mn} \text{sign}(m+z_3-m_0)$. 

The final step is to kill the unwanted factor $\text{sign}(m+z_3-m_0)$ in this commutator. We will of course use the freedom related to the choice of $m_0$. If $[x]$ denotes the integer part of $x$, we introduce the following functions \cite{Kallo1}
\begin{eqnarray}
\text{st}(x) & = & \begin{cases}
			\lbrack x \rbrack + 1   \quad \text{if} \quad x \geq 0 \\
			\lbrack x \rbrack \quad \quad \;\;\, \text{if} \quad x < 0
                   \end{cases}, \nonumber \\
m_0(x)       & = & \text{st}(x) - \frac{1}{2}, \nonumber \\
\zeta(x)     & = & x - m_0(x). \nonumber
\end{eqnarray}
These functions are presented in figure 1.
\begin{figure}[ht]
\begin{center}
\includegraphics{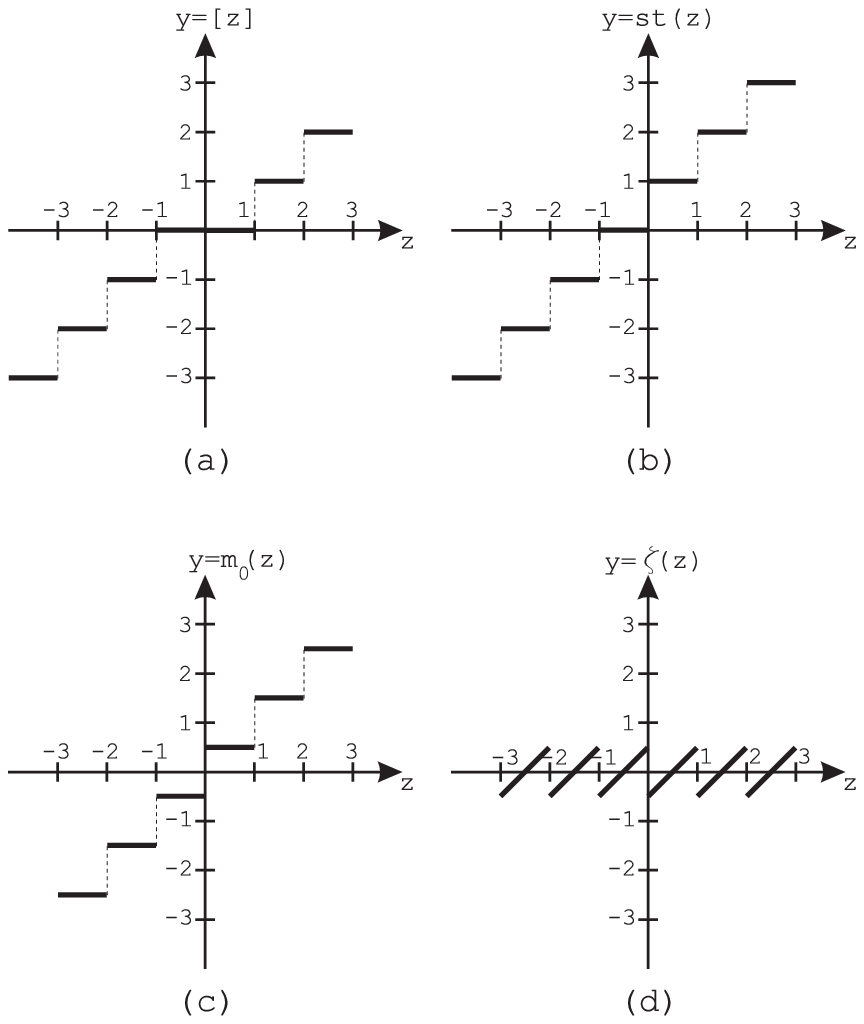}
\caption{{\em Functions (a) $[z]$, (b) $\rm{st}$$(z)$, (c) $m_0(z)$ and (d) $\zeta(z)$}.}
\end{center}
\end{figure}

It can easily be checked that they satisfy the following properties
\begin{eqnarray}\label{fctprop}
m_0(z+1)   & = & m_0(z) + 1, \nonumber \\
m_0(-z)    & = & - m_0(z),   \nonumber \\
\zeta(z+1) & = & \zeta(z),             \\
\zeta(-z)  & = & - \zeta(z), \nonumber \\
-\frac{1}{2} < & \zeta(z) & < \frac{1}{2}. \nonumber 
\end{eqnarray}
Defining
\begin{equation}
b_{m,2} = \tilde{A}_m, \quad\quad d_{m,2} = \tilde{A}_{-m} \quad \quad \text{for} \quad m\geq \frac{1}{2},
\end{equation}
we finally have
\begin{equation}\label{vpmq}
\vp_2 = \frac{e^{i\frac{m_0(z_3) \pi}{L} x^-}}{\sqrt{4\pi}} 
        \sum_{m=\frac{1}{2}}^\infty \left(
		u_{m,2} b_{m,2} e^{-i\frac{m \pi}{L} x^-} + v_{m,2} d_{m,2}^\dagger e^{i\frac{m \pi}{L} x^-}
				    \right)
\end{equation}
where\footnote{The last property in \eqref{fctprop} assure positivity of the arguments of the square roots.} $u_{m,2} = \frac{1}{\sqrt{m + \zeta(z_3)}}$ and $v_{m,2} = \frac{1}{\sqrt{m - \zeta(z_3)}}$.

Moreover, from
\[
\text{sign}(m+z-m_0) = \text{sign}(m-\zeta(z)) 
                     = \begin{cases}
			1 \;\;\: \quad \text{if} \quad m \geq \frac{1}{2} \\
			-1     \quad \text{if} \quad m \leq -\frac{1}{2}
		       \end{cases},
\]
a direct calculation shows that
\begin{equation}
[b_{m,2}, b_{n,2}^\dagger] = [d_{m,2}, d_{n,2}^\dagger] = \delta_{mn},
\end{equation}
while other commutators vanish.

Our last task in this section is to quantize the four remaining fields $\vp_4$, $\vp_5$, $\vp_6$, $\vp_7$. Equations \eqref{phicomp} teach us that we may quantize them by performing the replacements shown in Table 1.

\begin{table}[ht]
\begin{center}
\begin{tabular}{| c | c |}
\hline
Field & Functions \\
\hline
\hline
$\vp_1 \rightarrow \vp_5$ & $m_{0,1}=m_{0,2}\equiv m_0(z_3) \rightarrow m_{0,4}=m_{0,5}\equiv m_0(-\frac{z_3}{2}-z_8)$ \\
$\vp_2 \rightarrow \vp_4$ & $\zeta_1=\zeta_2\equiv\zeta(z_3) \rightarrow \zeta_4=\zeta_5\equiv\zeta(-\frac{z_3}{2}-z_8)$\\
\hline
$\vp_1 \rightarrow \vp_6$ & $m_{0,1}=m_{0,2}\equiv m_0(z_3) \rightarrow m_{0,6}=m_{0,7}\equiv m_0(-\frac{z_3}{2}+z_8)$ \\
$\vp_2 \rightarrow \vp_7$ & $\zeta_1=\zeta_2\equiv\zeta(z_3) \rightarrow \zeta_6=\zeta_7\equiv\zeta(-\frac{z_3}{2}+z_8)$\\
\hline
\end{tabular}
\end{center}
\caption{Replacements to perform for obtaining $\vp_4, \vp_5, \vp_6$ and $\vp_7$ from $\vp_1$ and $\vp_2$.}
\end{table}

Let us now summarize all informations obtained from quantization.

\begin{eqnarray*}
\vp_j & = & \frac{a_{0,j}}{\sqrt{4\pi}} 
          + \sum_{n=1}^\infty \frac{w_n}{\sqrt{4\pi}} \left(
	          a_{n,j} e^{-i\frac{n\pi}{L}x^-}
	        + a_{n,j}^\dagger e^{i\frac{n\pi}{L}x^-} \right)
\quad \quad \quad \quad \quad \quad \: \, j=3,8,\\
\vp_k & = & \frac{e^{im_{0,k}\frac{\pi}{L}x^-}}{\sqrt{4\pi}}
            \sum_{m=\frac{1}{2}}^\infty \left(
                  u_{m,k} b_{m,k} e^{-i\frac{m\pi}{L}x^-}
                + v_{m,k} d_{m,k}^\dagger e^{i\frac{m\pi}{L}x^-} \right)
\quad \quad \quad k=2,4,7, \\
\vp_1 & = & \vp_2^\dagger, \\
\vp_5 & = & \vp_4^\dagger, \\
\vp_6 & = & \vp_7^\dagger,
\end{eqnarray*}
with
\begin{eqnarray*}
w_n & = & \frac{1}{\sqrt{n}}, \\
u_{m,k} & = & \frac{1}{\sqrt{m+\zeta_k}}, \quad\quad\quad k=1,2,4,5,6,7,\\
v_{m,k} & = & \frac{1}{\sqrt{m-\zeta_k}}, \quad\quad\quad k=1,2,4,5,6,7,
\end{eqnarray*}

The only non-null commutators, except those involving the zero-modes, are
\begin{eqnarray*}
\left \lbrack a_{k,i}, a_{l,j}^\dagger \right \rbrack & = & \delta_{kl} \delta_{ij},\\
\left \lbrack b_{m,i}, b_{n,j}^\dagger \right \rbrack & = & \delta_{mn} \delta_{ij},\\
\left \lbrack d_{m,i}, d_{n,j}^\dagger \right \rbrack & = & \delta_{mn} \delta_{ij},
\end{eqnarray*}
where $k,l=1,2,\dots$ et $m,n=\frac{1}{2}, \frac{3}{2} \dots$.

\section{Back to gauge transformations}

\hspace{0.5cm} From the quantization results, it can be seen that we are still left with some discrete gauge transformations. These symmetries are those leaving $\zeta_2\equiv \zeta(z_3)$, $\zeta_4\equiv\zeta(-\frac{z_3}{2}-z_8)$ and $\zeta_7\equiv\zeta(-\frac{z_3}{2}+z_8)$ invariant. Given the property $\zeta(z+1)=\zeta(z)$, the most general form of such a transformation is
\begin{eqnarray}
z_3 & \rightarrow & z_3 + \alpha \nonumber \\[-3mm]
    &             & \quad \quad \quad \quad \quad \quad \quad \quad \quad \alpha, \beta \in \Z.
\\[-3mm]
z_8 & \rightarrow & z_8 + \frac{\alpha}{2} + \beta \nonumber
\end{eqnarray}
The property $m_0(z+1)=m_0(z)+1$ implies a single phase transformation for the adjoint field. The creation and destruction operators, and, as a consequence, the Fock vacuum,  are left invariant. These ``large gauge transformations'' correspond to Gribov copies \cite{Gribov1}.

\begin{figure}[ht]
\begin{center}
\includegraphics[angle=270,scale=0.26]{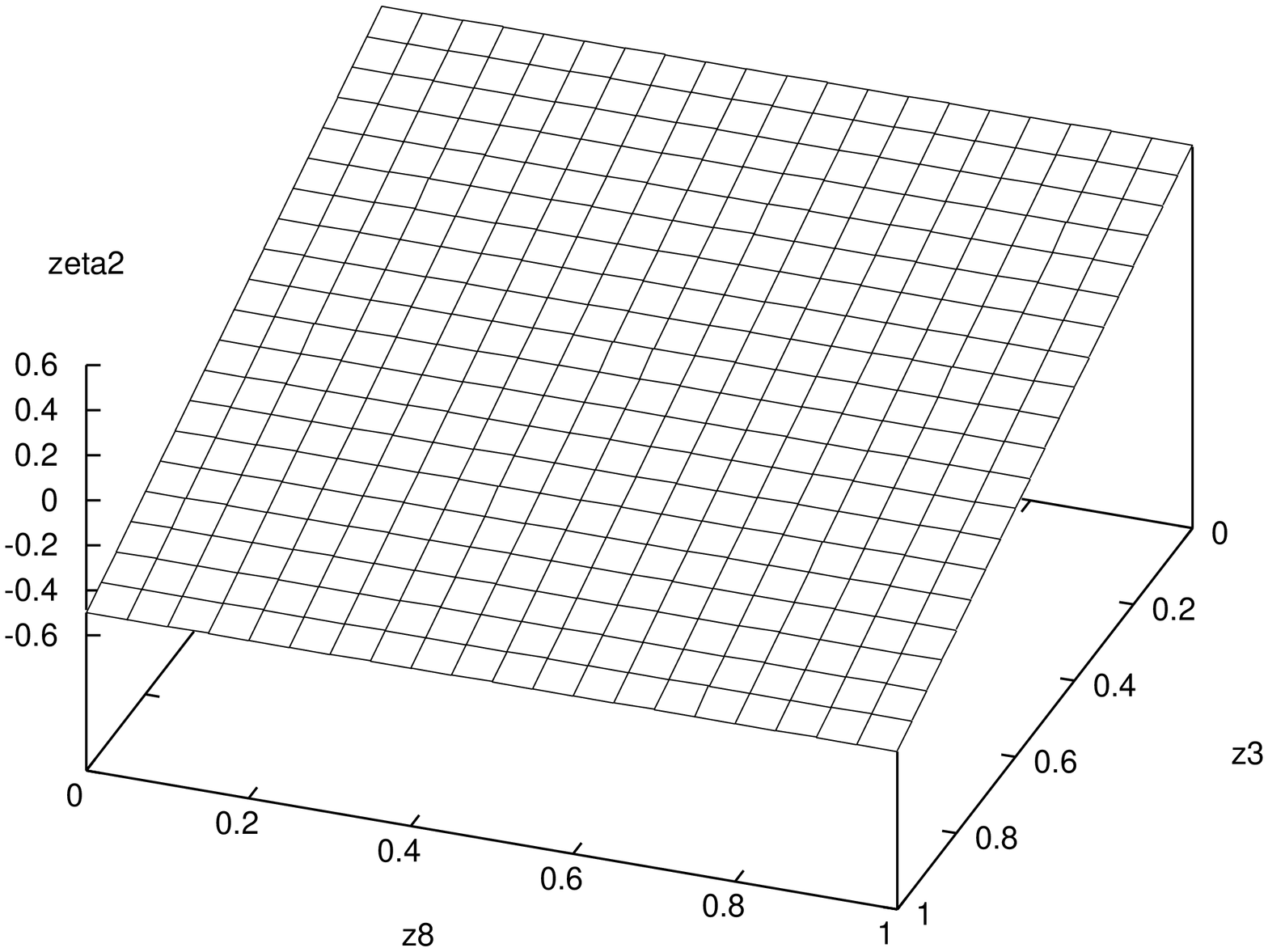}\hspace{-1.5cm}
\includegraphics[angle=270,scale=0.26]{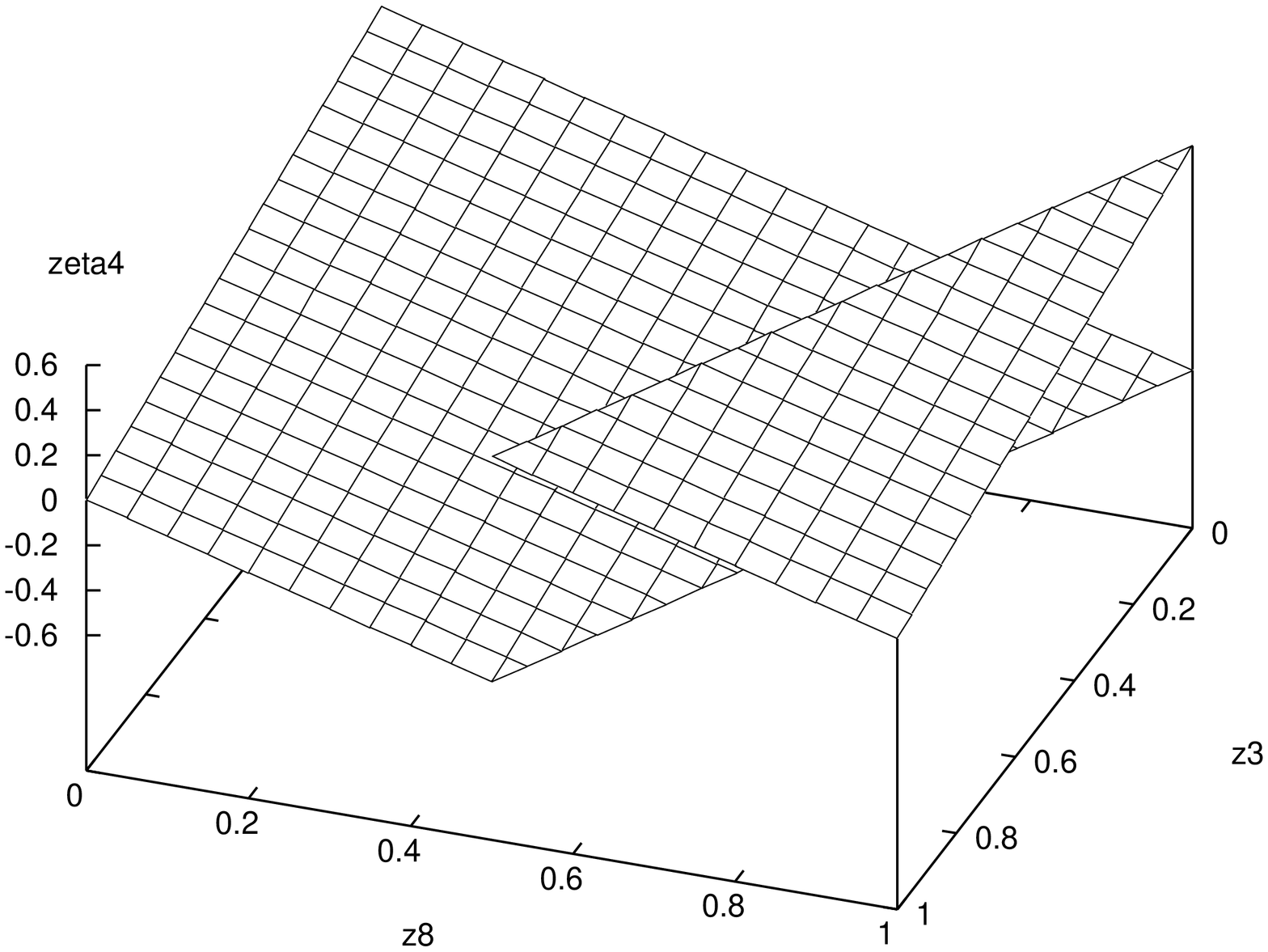}\hspace{-1.5cm}
\includegraphics[angle=270,scale=0.26]{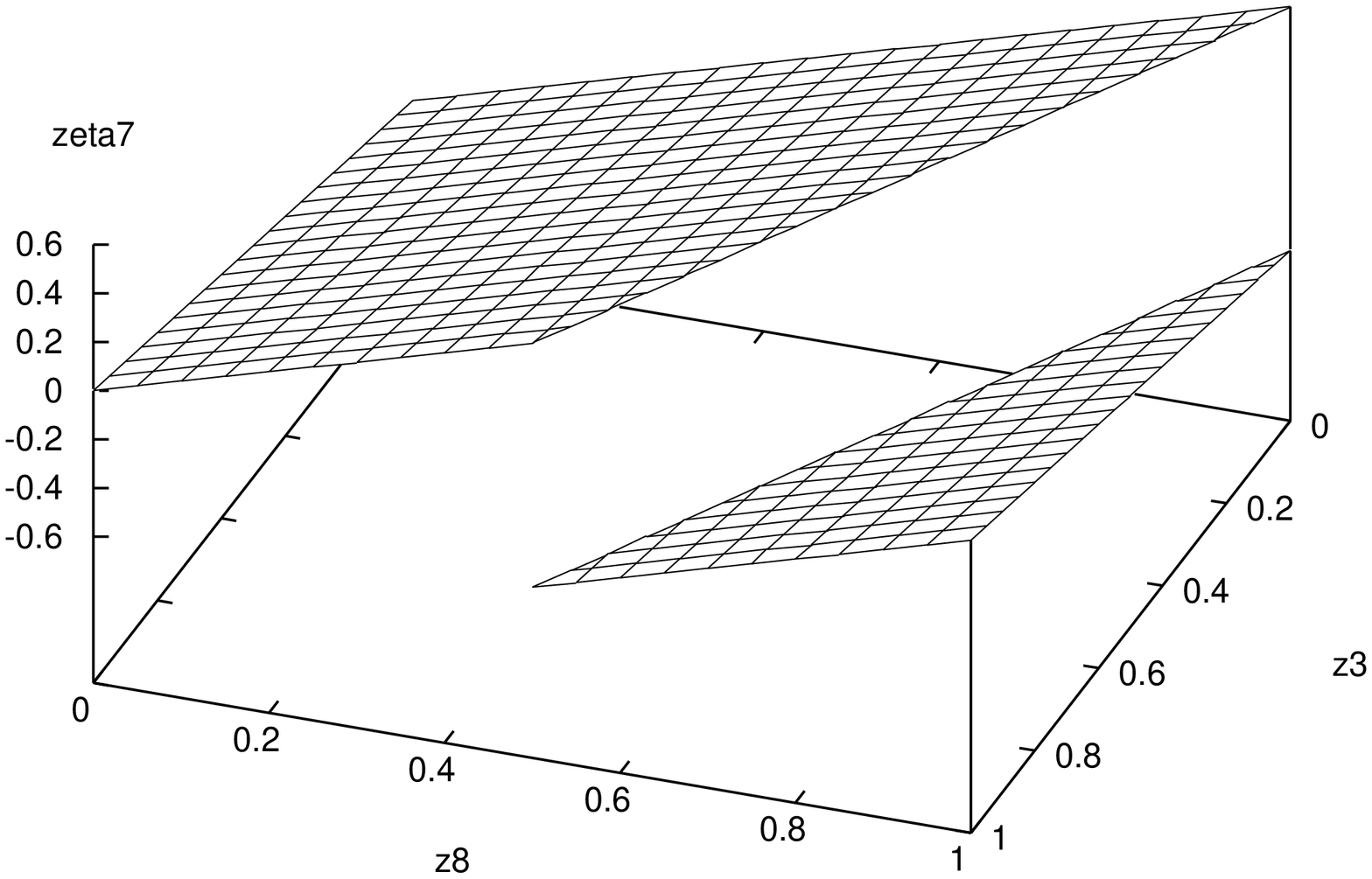}
\caption{{\em Graph for $\zeta_2$,  $\zeta_4$ and  $\zeta_7$ as functions of $z_3$ et $z_8$}.}
\end{center}
\end{figure}

Roughly speaking, this situation allows us to choose a fundamental domain \cite{domain}, for $z_3$ and $z_8$, in two different ways :
\begin{rmenum}
\item we take $0 \leq z_3 \leq 1$ and $0 \leq z_8 \leq 1$. The domain is simple but $\zeta_4$ and $\zeta_7$ present discontinuities as shown in figure 2.
\item we choose $0 \leq z_3 \leq 1$ and $z_3-\frac{1}{2} \leq z_8 \leq z_3+\frac{1}{2}$. This choice is less ``natural'' but removes all the discontinuities.
\end{rmenum}

\section{Gauss law and the matter current}

\hspace{0.5cm} We are now in good position for solving the Gauss law \eqref{eqnA}. Let's start by developing the matter current in Fourier series as follows\footnote{For simplicity, we shall omit the $M$ index for the matter current.}
\begin{eqnarray}
J_k^+(x)& = & -\frac{1}{4L} \sum_{n\in \Z} J_k(n) e^{-i\frac{n\pi}{L}x^-} 
     \quad \quad \quad \quad \quad \quad \;\:\: k=3,8, \nonumber \\
J_k^+(x)& = & -\frac{1}{4L} e^{im_{0,k}\frac{\pi}{L}x^-} 
    \sum_{m\in \HZ} J_k(m) e^{-i\frac{m\pi}{L}x^-} 
     \quad\quad\;\; k=2,4,7,\\
J_k^+(x)& = & -\frac{1}{4L} e^{-im_{0,k}\frac{\pi}{L}x^-} 
    \sum_{m\in \HZ} J_k(m) e^{-i\frac{m\pi}{L}x^-} \quad \quad k=1,5,6.
\nonumber
\end{eqnarray}

Using the definition of $\J$
\[
\J^+ = -i [\PHI, {\boldsymbol \pi}],
\]
and the structure constants in our $SU(3)$ basis, the $\J$ components can be written as a functions of the fields $\vp_i$ and $\pi_i$. We obtain, in a straightforward way,
\begin{eqnarray}
J^+_3 & = & -i\left( \vp_1\pi_2 - \vp_2\pi_1 + \frac{1}{2}\vp_4\pi_5 - \frac{1}{2}\vp_5\pi_4
                   + \frac{1}{2}\vp_6\pi_7 - \frac{1}{2}\vp_7\pi_6 \right)_s, \nonumber\\
J^+_8 & = & -\frac{3i}{4} \left( \vp_4\pi_5 - \vp_5\pi_4 + \vp_6\pi_7 - \vp_7\pi_6 \right)_s, \nonumber\\
J^+_2 & = &  i \left( \vp_3\pi_2 - \vp_2\pi_3 + \frac{1}{\sqrt{2}}\vp_5\pi_6 - \frac{1}{\sqrt{2}}\vp_6\pi_5 \right)_s,
            \nonumber\\
J^+_4 & = & -i \left( \frac{1}{2}\vp_3\pi_4 - \frac{1}{2}\vp_4\pi_3 + \vp_8\pi_4 - \vp_4\pi_8
                    + \frac{1}{\sqrt{2}}\vp_1\pi_6 - \frac{1}{\sqrt{2}}\vp_6\pi_1 \right)_s, \nonumber\\
J^+_7 & = & -i \left( \frac{1}{2}\vp_3\pi_7 - \frac{1}{2}\vp_7\pi_3 - \vp_8\pi_7 + \vp_7\pi_8
                    - \frac{1}{\sqrt{2}}\vp_1\pi_5 + \frac{1}{\sqrt{2}}\vp_5\pi_1 \right)_s
\end{eqnarray}
where the $s$ index means that these expressions are symmetrized in order to preserve hermiticity.
We have similar expressions for $J^+_1$, $J^+_5$ and $J^+_6$.
Inserting the complete Fock expansions, found in the previous section, into these results will give the matter current as function of creation and destruction operators. One can easily understand that the final result is quite cumbersome. We send the interested reader to the appendix at the end of this paper to have an overview of the explicit development.

Recall that the matter current can be treated as a source for the $\A$ field as shown by the Gauss law \eqref{eqnA}. We can extract $\A$ from this Gauss equation. Formally, we may write
\begin{equation}\label{solGauss0}
\A = -g \frac{1}{\der_-^2} \J^+.
\end{equation}
Unicity of the solution is ensured by periodicity conditions and by the gauge conditions $\zm{A_3}=\zm{A_8}=0$. An important result can already be derived from equation \eqref{eqnA}. Projecting \eqref{eqnA} onto $\lambda^3$ and $\lambda^8$ gives
\begin{equation}
\partial_-^2 A_3 = -g J_3^+ \quad \quad \text{and} \quad \quad \partial_-^2 A_8 = -g J_8^+,
\end{equation}
which can be directly verified by using the second gauge condition and the fact that $\lambda^3$ and $\lambda^8$ are commuting. Taking zero-modes on both sides of these equations allows us to write
\[
I_3\equiv\zm{J_3^+} = 0 \quad \quad \text{and} \quad \quad Y\equiv\zm{J_8^+} = 0.
\]
These constraints can be realized by imposing that the physical states satisfy
\begin{equation}\label{cdtphys}
\boxed{I_3\ket{\text{phys}}=0} \quad  \quad \text{and} \quad \quad \boxed{Y\ket{\text{phys}}=0.}
\end{equation}
The operators $I_3$ et $Y$ will be respectively called {\bf isospin} and {\bf hypercharge} by analogy with the quark-parton model. Moreover, we can continue this analogy a little bit further. Writing
\[
(\der_- \PHI)_a = (\partial_- + ig\eta_a v_3 + ig \xi_a v_8) \vp_a,
\]
we can map all the $\vp_a$ on two weight diagrams associated with the $3$ and $\bar{3}$ representations in $SU(3)$.

\begin{figure}[ht]
\begin{center}
\includegraphics{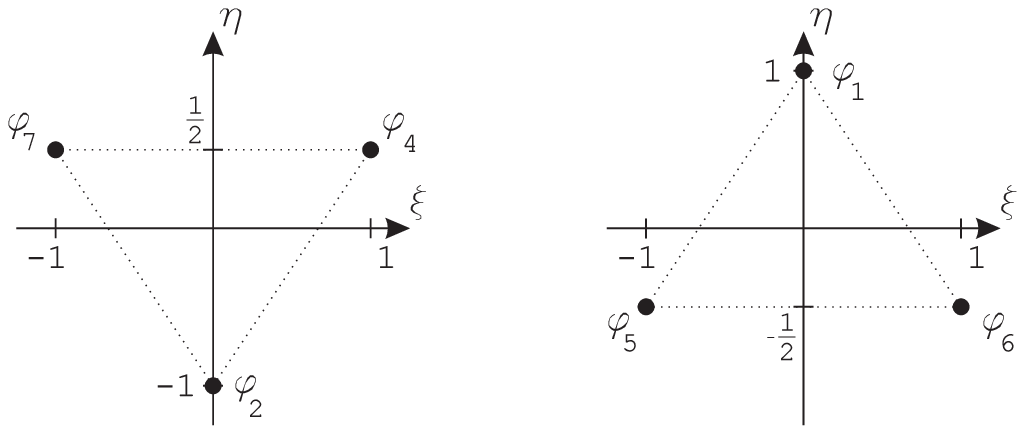}
\caption{{\em $SU(3)$ fundamental representations and adjoint field}.}
\end{center}
\end{figure}

Expanding $I_3$ and $Y$ over creation and destruction operators gives after some algebra
\begin{eqnarray}
I_3 & \sim & \sum_{m=\frac{1}{2}}^\infty \left(b_{m,2}^\dagger b_{m,2} - d_{m,2}^\dagger d_{m,2} \right)
                            -\frac{1}{2} \left(b_{m,4}^\dagger b_{m,4} - d_{m,4}^\dagger d_{m,4} \right)
                            -\frac{1}{2} \left(b_{m,7}^\dagger b_{m,7} - d_{m,7}^\dagger d_{m,7} \right),
             \nonumber \\
Y   & \sim & \sum_{m=\frac{1}{2}}^\infty - \left(b_{m,4}^\dagger b_{m,4} - d_{m,4}^\dagger d_{m,4} \right)
                                         + \left(b_{m,7}^\dagger b_{m,7} - d_{m,7}^\dagger d_{m,7} \right).
\end{eqnarray}

We can interpret the former relation by associating quantum numbers to the adjoint field components and the ladder operators. These quantum numbers are given in table 2.

\begin{table}
\begin{center}
\begin{tabular}{|c||c|c||c|c|}
\hline
        &               &       & $I_3$          & $Y$  \\
\hline
\hline
$\vp_1$ & $b_2^\dagger$ & $d_2$ & $1$            & $0$  \\
\hline
$\vp_2$ & $d_2^\dagger$ & $b_2$ & $-1$           & $0$  \\
\hline
$\vp_4$ & $d_4^\dagger$ & $b_4$ & $\frac{1}{2}$  & $1$  \\
\hline
$\vp_5$ & $b_4^\dagger$ & $d_4$ & $-\frac{1}{2}$ & $-1$ \\
\hline
$\vp_6$ & $b_7^\dagger$ & $d_7$ & $-\frac{1}{2}$ & $1$  \\
\hline
$\vp_7$ & $d_7^\dagger$ & $b_7$ & $\frac{1}{2}$  & $-1$ \\
\hline
\end{tabular}
\end{center}
\caption{{\em Quantum numbers associated with the adjoint field}.}
\end{table}
It can be directly checked that this is consistent with the previous weight diagrams.

\section{Constrained zero-modes}

\hspace{0.5cm} We are now able to deduce the constraints on the two zero-modes $a_{0,3}$ and $a_{0,8}$. Our starting point will be equation \eqref{mvtPHI}. Having in mind the $SU(2)$ model, we add a mass term to this equation which becomes
\begin{equation}
\der^\alpha \der_\alpha \PHI + \mu_0^2 \PHI = 0,
\end{equation}
remembering that we are working with a dimensionally reduced $1+1$ dimensions model.
The introduction of a mass term may be justified by an unavoidable renormalization. Although we won't reach that point in this paper, we always have the freedom to set the renormalized mass to zero at the very end. The mass term can be obtained by subtracting $\tr(\mu_0^2\PHI^2)$ to the Lagrangian.

We will limit our development to the establishment of the constraints without entering into their resolution. The trick is to expand the relation
\begin{equation}\label{contr0}
\boxed{
\tr\left(\Zm{ \left( \der^\alpha \der_\alpha \PHI +\mu_0^2 \PHI \right) \la^j }\right) = 0,} \quad 
\quad \quad j=3,8,
\end{equation}
which is a direct consequence of the previous equation. Using the property that $\Zm{\partial_- f}=0$, true for every periodic function $f$, and the structure constants, one can bring back \eqref{contr0} to the form
\[
ig \tr\left( \Zm{ [\A, \der_-\PHI] \la^j } \right) + \mu_0^2 \tr\left( \PHI \la^j \right) = 0, 
\quad \quad \quad j=3,8.
\]
Expanding these relations on the chosen $su(3)$ basis and using the Gauss law \eqref{eqnA}, two constraints are obtained
\begin{eqnarray}
-i \left\langle 
               \frac{1}{2} \vp_5 \frac{1}{\partial_-+\frac{ig}{2}v_3+igv_8} J^+_4 
             - \frac{1}{2} \vp_4 \frac{1}{\partial_--\frac{ig}{2}v_3-igv_8} J^+_5
\right. & & \nonumber \\
	     + \frac{1}{2} \vp_6 \frac{1}{\partial_-+\frac{ig}{2}v_3-igv_8} J^+_7 
             - \frac{1}{2} \vp_7 \frac{1}{\partial_--\frac{ig}{2}v_3+igv_8} J^+_6 
        & & \label{a3contr} \\ \left. 
           + \vp_2 \frac{1}{\partial_-+igv_3} J^+_1 - \vp_1 \frac{1}{\partial_--igv_3} J^+_2 
\right\rangle_{0,s}
 & = & \frac{\mu_0^2}{\sqrt{4\pi}g^2} a_{0,3}, \nonumber\\
-i \left\langle \vp_5 \frac{1}{\partial_-+\frac{ig}{2}v_3+igv_8} J^+_4 
             - \vp_4 \frac{1}{\partial_--\frac{ig}{2}v_3-igv_8} J^+_5 \right.& & \nonumber \\
[-3mm] & & \label{a8contr} \\
      \left. - \vp_6 \frac{1}{\partial_-+\frac{ig}{2}v_3-igv_8} J^+_7 
             + \vp_7 \frac{1}{\partial_--\frac{ig}{2}v_3+igv_8} J^+_6 \right\rangle_{0,s}
 & = & \frac{\mu_0^2}{\sqrt{4\pi}g^2} \frac{4}{3} a_{0,8}. \nonumber
\end{eqnarray}

Again, it is, of course, possible to write the constraints in terms of creation and destruction operators. 

Following \cite{Kallo2}, it is also possible to translate the constraints into diagrams by introducing vertex and propagators. 

We won't go into such a tedious task in the present paper.
If we make the assumption that we can renormalize the constraints and if we bring back the constrained zero-modes to the 3-particles sector, we should expect a unique solution for these zero-modes. Unicity of the solution is enforced by the fact that the constraints are linear in the zero-modes. That should mean that the $SU(3)$ symmetry is not broken. This situation differs from the $\phi^4$ theory where we have a cubic constraint giving rise to spontaneous symmetry breaking.
These developments are defered to a future paper.

\section{Vacuum and dynamical zero-modes}

\hspace{0.5cm} In this section, we finally reach the aim of this paper, which is to study the vacuum degeneracy. It is a well known result that constrained zero-modes are related to symmetry breaking while dynamical zero-modes are related to vacuum degeneracy. In order to emphasize the contribution of the two dynamical zero-modes $z_3$ and $z_8$ on the vacuum of our model, we write it as the product of a Fock part and a zero-mode part. If $\ket{\Omega}$ is the vacuum and $\ket{0}$ is the Fock vacuum, which means that it is annihilated by all destruction operators, we have
\begin{equation}
\ket{\Omega} = \ket{0} \otimes \ket{\Psi_0(z_3,z_8)},
\end{equation}
where $\ket{\Psi_0(z_3,z_8)}$ depend only on the two dynamical zero-modes.

We are going to write down a Schr\"{o}dinger equation for $\ket{\Psi_0}$. Developing the Hamiltonian from equation \eqref{hamilt}, leads to
\begin{eqnarray*}
H & = & -4 \frac{\partial^2}{\partial z_3^2} - 3 \frac{\partial^2}{\partial z_8^2} \\
  & + & \sum_{k=0}^\infty w_{k,3}^4 \left( J_3(k)J_3^\dagger(k) + J_3^\dagger(k)J_3(k) \right)
          + \frac{3}{4} w_{k,8}^4 \left( J_8(k)J_8^\dagger(k) + J_8^\dagger(k)J_8(k) \right) \\
  & + & \sum_{k=\frac{1}{2}}^\infty v_{k,2}^4 \left( J_1(k)J_1^\dagger(k) + J_1^\dagger(k)J_1(k) \right)
                                  + u_{k,2}^4 \left( J_2(k)J_2^\dagger(k) + J_2^\dagger(k)J_2(k) \right) \\
  & + & \begin{pmatrix} 1 \rightarrow 5 \\ 2 \rightarrow 4 \end{pmatrix}  
       +\begin{pmatrix} 1 \rightarrow 6 \\ 2 \rightarrow 7 \end{pmatrix}.
\end{eqnarray*}
Defining $H_0 = \scalop{0}{H}{0}$, we have to solve the equation
\begin{equation}\label{ScSU3}
H_0 \ket{\Psi_0(z_3,z_8)} = E_0 \ket{\Psi_0(z_3,z_8)}.
\end{equation}

Given that a complete treatment of this equation needs a solution for the constrained zero-modes, we shall restrict ourselves to the normal-mode sector by simply neglecting the constrained zero-mode contribution. As shown by Kalloniatis in \cite{Kallo2}, this contribution is mich smaller than those coming from the other terms. A model including the constrained zero-mode part is left for future work. In such a case, we can successively find $\J$ and $H$ in terms of the creation and destruction operators and then calculate $H_0$. After some straightforward but tedious calculations, we get
\begin{equation}\label{decomp}
H_0 = -4 \frac{\partial^2}{\partial z_3^2} - 3 \frac{\partial^2}{\partial z_8^2} + V_0,
\end{equation}

\begin{eqnarray*}
V_0 & = & \sum_{k,m=\frac{1}{2}}^\infty \frac{1}{(k+m)^2}
             \left[\frac{(k-m-2\zeta_2)^2}{(m+\zeta_2)(k-\zeta_2)}
                   + (\zeta_2 \rightarrow \zeta_4) + (\zeta_2 \rightarrow \zeta_7) \right] \\
    & + & \sum_{m=\frac{1}{2}}^\infty \sum_{k=m+1}^\infty \frac{1}{k-m} \left[
             \left(\frac{k-2m+\zeta_2}{(k-\zeta_2)^2(m-\zeta_2)} + \frac{k-2m-\zeta_2}{(k+\zeta_2)^2(m+\zeta_2)} \right)
           + (\zeta_2 \rightarrow \zeta_4) + (\zeta_2 \rightarrow \zeta_7) \right] \\
    & + & \left. \sum_{m,n=\frac{1}{2}}^\infty \right[
          \left( \frac{(m-n+\zeta_7-\zeta_4)^2}{(m+n+\zeta_2+M_0)^2(m-\zeta_4)(n-\zeta_7)}
               + \frac{(m-n-\zeta_7+\zeta_4)^2}{(m+n-\zeta_2-M_0)^2(m+\zeta_4)(n+\zeta_7)} \right) \\
    &   & \quad \quad \left. + \begin{pmatrix} \zeta_2 \rightarrow \zeta_4 \\
                                               \zeta_4 \rightarrow \zeta_7 \\
                                               \zeta_7 \rightarrow \zeta_2 \end{pmatrix}
                             + \begin{pmatrix} \zeta_2 \rightarrow \zeta_7 \\
                                               \zeta_4 \rightarrow \zeta_2 \\
                                               \zeta_7 \rightarrow \zeta_4 \end{pmatrix} \right].
\end{eqnarray*}
In this expression, we have developed the $u_{m,k}$, $v_{m,k}$ and $w_{m,k}$ coefficients in terms of $m$, $l$ and $\zeta_k$, and introduced $M_0(z_3, z_8) = m_{0,2}+m_{0,4}+m_{0,7}$.

In order to solve \eqref{ScSU3}, we shall work with the fundamental domain $0 \leq z_3 \leq 1$, $z_3-\frac{1}{2} \leq z_8 \leq z_3+\frac{1}{2}$. Performing the change of variables
\[
\begin{cases}
u = z_3 \\
v = z_8 + \frac{1}{2} z_3
\end{cases}, \quad \quad \quad (u,v) \in [0,1] \times [0,1],
\]
equation \eqref{ScSU3} becomes
\begin{equation}\label{fineq}
\left[ -4\left( \partial_u^2 + \partial_u \partial_v + \partial_v^2 \right) + V_0(u,v) \right] \Psi_0(u,v) = E_0 \Psi_0(u,v).
\end{equation}
Numerically, we are able to calculate the potential $V_0(u,v)$ and thus to solve this equation. The result is shown in figure 4.

\begin{figure}[ht]
\begin{center}
\includegraphics[scale=0.33,angle=270]{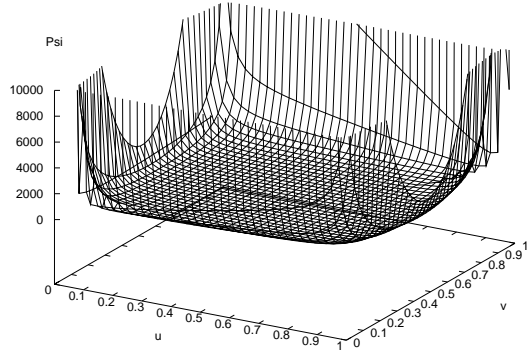}\\
\caption{{\em Potential $V_0(u,v)$}.}
\end{center}
\end{figure}

\begin{figure}
\begin{center}
\includegraphics[scale=0.33,angle=270]{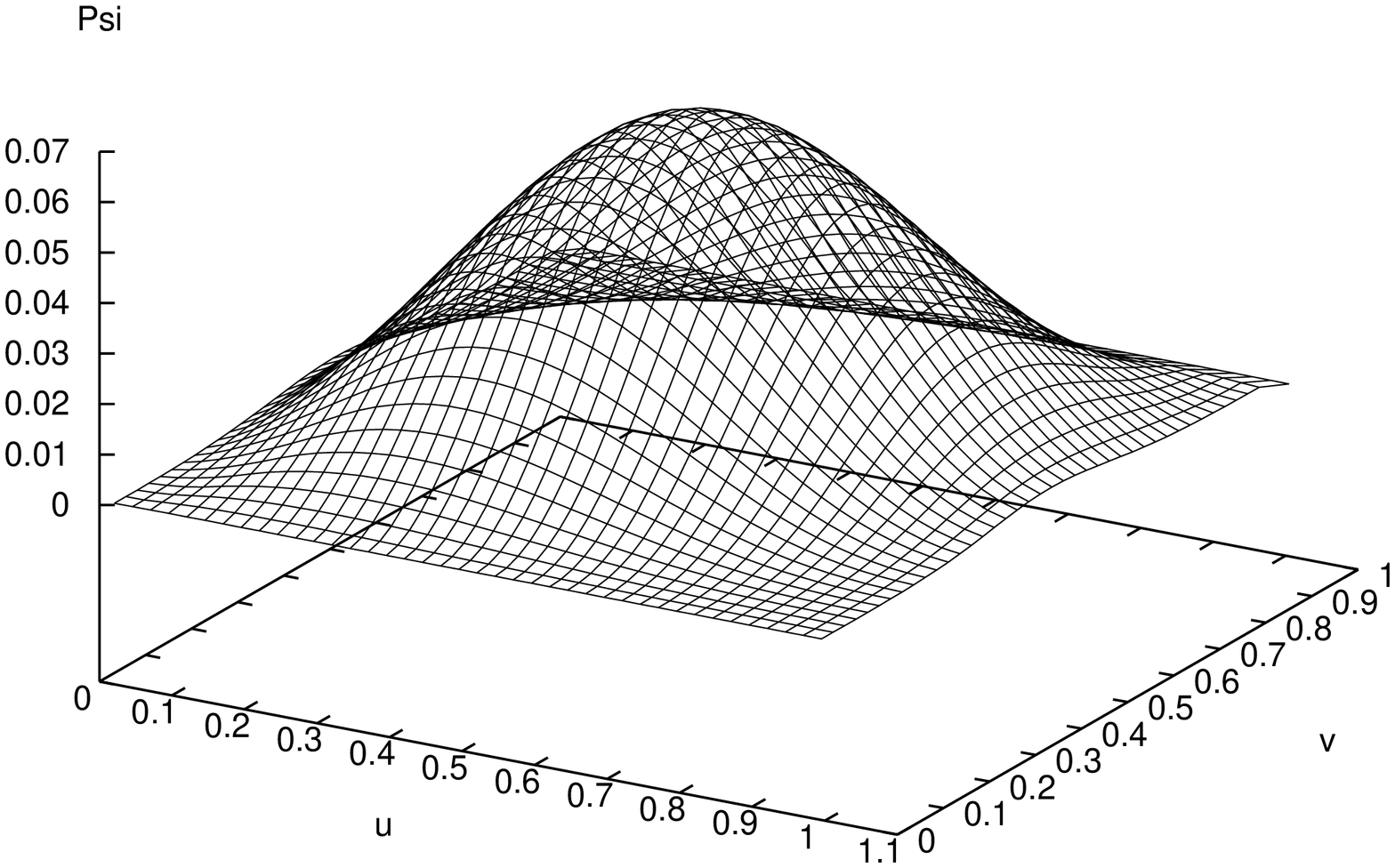}\hspace{-0.9cm}
\includegraphics[scale=0.33,angle=270]{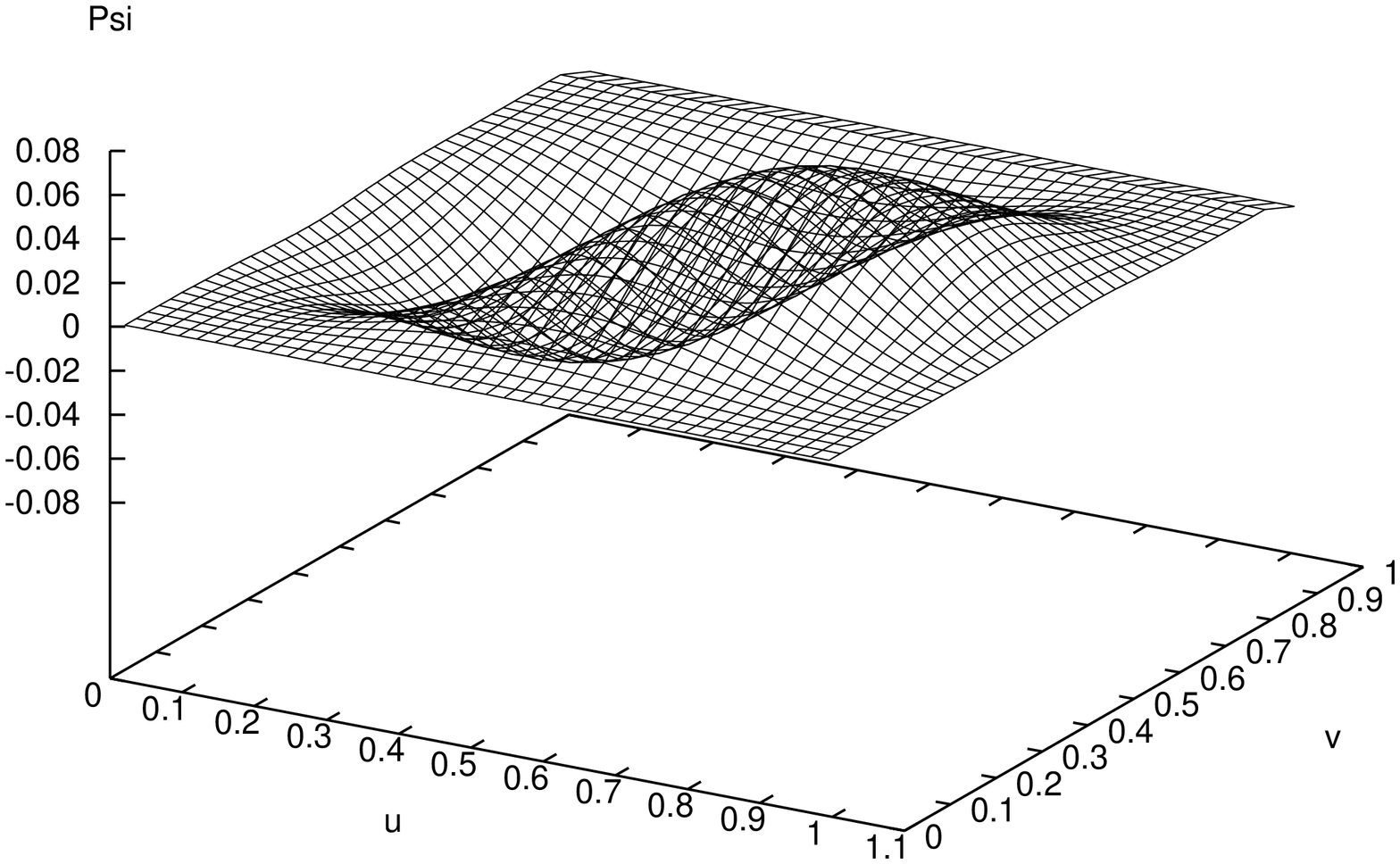}\\
\caption{{\em Fundamental state $\Psi_0(u,v)$ and first excited state $\Psi_1(u,v)$}.}
\end{center}
\end{figure}

One can see that $V_0$ presents discontinuities at the boundaries of the domain\footnote{Working with a fundamental domain in which discontinuities are at the boundaries is more explicit and is easier from a numerical point of view.}. Given the form of the potential $V_0$, me may assimilate it to a two dimensional square well and solve. It has a minimum at $u=v=0.5$.

Unfortunately the crossed term $-4 \partial_u\partial_v \Psi(u,v)$ in equation \eqref{fineq} distinguishes our eigenvalue problem from the traditional ``square well'' one. No analytic solution of this equation has been found.

Unicity of the minimum of $V_0$ suggests us that the vacuum is non-degenerate. A numerical diagonalization of equation \eqref{fineq} allows us to obtain the wave functions $\ket{\Psi}$ and their energy levels. The ground state and the first excited state are given in figure 5. The numerical results for the energy levels confirm that  
\begin{center}
{\em if we neglect the constrained zero-modes, vacuum is non-degenerate.}
\end{center}


Setting $V(u=\frac{1}{2}, v=\frac{1}{2})=0$, we have $E_0 \approx 138.3$. This can be seen as a zero-point energy. More precisely, the physical zero-point energy is given by $\frac{g^2 L}{4\pi^2}E_0$. It is the minimum energy of any physical system described by Lagrangian \eqref{TheL}.

Note that vacuum degeneracy is not definitively impossible. Renormalization and constraint zero-modes contribution should lead to vacuum degeneracy. As shown in \cite{SUSY}, a supersymmetric calculation of this model leads to a SUSY potential in \eqref{decomp} cancelling parts of $V_0$ and giving rise to vacuum degeneracy.

\section{Conclusions}

\hspace{0.5cm} In this paper, we have seen that, under some assumptions,
we were able to study the impact of zero-modes on a non abelian theory
and especially on its vacuum. 

Like in many light-cone quantization models, we have seen the simplifications coming from the use of 
creation and annihilation operators.

The aim of this paper was to study vacuum degeneracy. We thus neglected constrained zero-modes and turned to dynamical zero-modes. We arrived at a Schr\"{o}dinger-like
equation, neglecting contribution from the constrained zero-modes.
Numerical resolution of this equation leads to the conclusion that, under our
approximations, the $SU(3)$ light-cone vacuum is non-degenerate.

This model is a first step in our quest to reach the QCD case. The remaining work will involve the elimination of the approximations of this model : fields independent of the
transverse variable and solution for the constrained zero-modes. If fields are dependant of the tranverse variable, we have a full $2+1$ dimensional model and we can't anymore dimensionally reduce it to $1+1$ dimensions. A solution of the constraints, even if not exact, should have an impact on the vacuum degeneracy, modifying the potential in the Schr\"{o}dinger equation. 

We then may hope to be able to establish a model in $3+1$ dimensions. Practically, this introduces two adjoint fields and coupling between these. Such a generalization should again lead to more tedious calculations, especially during the quantization of adjoint fields.
 
The final step is to add quark fields. A $2+1$ dimensional $SU(2)$ model coupled to a fermion field has been introduced by M. Tachibana \cite{tachi}. This is, in fact, a difficult application of the Dirac-Bergman quantization method.

All these points are thus far away from being trivial. Some new techniques or approximations are certainly needed if we want to reach finally the full QCD case.

\vspace{1.5cm}

\begin{center}
{\large{ACKNOWLEDGMENTS}}\vspace{0.8cm}
\end{center}
This work was supported by National Fund for Scientific Research, Belgium. I would like to thank A. Burnel for guiding me during both development and writing of this paper.

\newpage

\begin{appendix}

\section{$SU(3)$ conventions}\label{appSU3}

\hspace{0.5cm} This appendix contains a description of the conventions used for the gauge group $SU(3)$.
The generators basis we used can be defined in $SU(N)$. It is a direct generalization of the basis associated with the light-cones studies of the $SU(2)$ gauge group. We define the matrices $\lambda_{ij}$ for $1\leq i,j \leq N$ and $(i,j) \neq (N,N)$ as
\begin{equation}
\left(\lambda^{ij}\right)_{ab} = \begin{cases}
\frac{1}{i+1} \sum_{k=1}^i \delta_{ak}\delta_{bk} - i\delta_{a,i+1}\delta_{b,i+1}
                \quad \, \text{if} \quad  i=j\\
\frac{1}{\sqrt{2}} \delta_{ai}\delta_{bj}
                \quad\quad\quad\quad\quad\quad\quad\quad\quad\quad \: \, \text{otherwise}
\end{cases}.
\end{equation}

These $N^2-1$ matrices are traceless and linearly independent.

Particularly, for $N=3$ they can be written as
\begin{equation}
\begin{matrix}
\lambda^{11}=\frac{1}{2}\begin{pmatrix}
1 & .  & .  \\
. & -1 & .  \\
. & .  & .
\end{pmatrix}, &
\lambda^{12}=\frac{1}{\sqrt{2}}\begin{pmatrix}
. & 1  & . \\
. & .  & . \\
. & .  & .
\end{pmatrix}, &
\lambda^{13}=\frac{1}{\sqrt{2}}\begin{pmatrix}
. & .  & 1 \\
. & .  & . \\
. & .  & . 
\end{pmatrix} \\
\lambda^{21}=\frac{1}{\sqrt{2}}\begin{pmatrix}
. & .  & .  \\
1 & .  & .  \\
. & .  & .
\end{pmatrix}, &
\lambda^{22}=\frac{1}{3}\begin{pmatrix}
1 & .  & . \\
. & 1  & . \\
. & .  & -2
\end{pmatrix}, &
\lambda^{23}=\frac{1}{\sqrt{2}}\begin{pmatrix}
. & .  & . \\
. & .  & 1 \\
. & .  & .
\end{pmatrix}, \\
\lambda^{31}=\frac{1}{\sqrt{2}}\begin{pmatrix}
. & .  & .  \\
. & .  & .  \\
1 & .  & .
\end{pmatrix}, &
\lambda^{32}=\frac{1}{\sqrt{2}}\begin{pmatrix}
. & .  & . \\
. & .  & . \\
. & 1  & .
\end{pmatrix}. & \,
\end{matrix}
\end{equation}

In analogy with the Gell-Mann matrices and in order to simplify index manipulations, we set
$\lambda^{1}\equiv \lambda^{12}$, $\lambda^{2}\equiv \lambda^{21}$, $\lambda^{3}\equiv \lambda^{11}$, 
$\lambda^{4}\equiv \lambda^{13}$, $\lambda^{5}\equiv \lambda^{31}$, $\lambda^{6}\equiv \lambda^{23}$, 
$\lambda^{7}\equiv \lambda^{32}$, $\lambda^{8}\equiv \lambda^{22}$.

With this convention, we can calculate the structure constants and $SU(3)$ metric. Taking as a convention that 
\begin{equation}
\left\lbrack \lambda^i, \lambda^j \right\rbrack = {f^{ij}}_k \lambda^k, \quad \text{and} \quad
\tr(\lambda_i \lambda^j) = 2 \delta_i^j,
\end{equation}
we have
\begin{eqnarray} 
f^{ijk}        & = & 2 \, \tr([\lambda^i,\lambda^j]\lambda^k), \\
{\cal{G}}^{ij} & = & 2 \, \tr(\lambda^i \lambda^j).
\end{eqnarray}
From the relation ${\cal{G}}^{ij} {\cal{G}}_{jk} = \delta_i^k$, we deduce ${\cal{G}}_{jk}$ by simply inverting ${\cal{G}}^{ij}$.

With the matrices $\lambda_j$ introduced before, we find

\begin{equation}
{\cal{G}}^{ij} = \begin{pmatrix}
. & 1 & . & . & . & . & . & .\\
1 & . & . & . & . & . & . & .\\
. & . & 1 & . & . & . & . & .\\
. & . & . & . & 1 & . & . & .\\
. & . & . & 1 & . & . & . & .\\
. & . & . & . & . & . & 1 & .\\
. & . & . & . & . & 1 & . & .\\
. & . & . & . & . & . & . & \frac{4}{3}
\end{pmatrix}.
\end{equation}

The non-vanishing structure constants are\footnote{We have taken into account the antisymmetry of the structure constants under permutation of the first two indices.}
\begin{center}
\begin{tabular*}{9.24cm}{| c | c | c | c || c | c | c | c || c | c | c | c |}
\hline
$i$     & $j$     & $k$     & $f^{ij}_{\;\;\;k}$        &
$i$     & $j$     & $k$     & $f^{ij}_{\;\;\;k}$        &
$i$     & $j$     & $k$     & $f^{ij}_{\;\;\;k}$        \\
\hline
\hline
1       & 2       & 3       & 1                         &
{\bf 3} & {\bf 4} & {\bf 4} & $\mathbf{\frac{1}{2}}$    &
6       & 7       & 3       & $\frac{-1}{2}$            \\

1       & 6       & 4       & $\frac{1}{\sqrt{2}}$      &
{\bf 3} & {\bf 6} & {\bf 6} & $\mathbf{\frac{-1}{2}}$   &
6       & 7       & 8       & $\frac{-3}{4}$            \\

2       & 4       & 6       & $\frac{1}{\sqrt{2}}$      &
{\bf 3} & {\bf 7} & {\bf 7} & $\mathbf{\frac{1}{2}}$    &
{\bf 8} & {\bf 4} & {\bf 4} & {\bf 1}                   \\

2       & 7       & 5       & $\frac{-1}{\sqrt{2}}$     &
4       & 5       & 3       & $\frac{1}{2}$             &
{\bf 8} & {\bf 5} & {\bf 5} & {\bf -1}                  \\

{\bf 3} & {\bf 1} & {\bf 1} & {\bf 1}                   &
4       & 5       & 8       & $\frac{3}{4}$             &
{\bf 8} & {\bf 6} & {\bf 6} & {\bf 1}                   \\

{\bf 3} & {\bf 2} & {\bf 2} & {\bf -1}                  &
4       & 7       & 1       & $\frac{1}{\sqrt{2}}$      &
{\bf 8} & {\bf 7} & {\bf 7} & {\bf -1}                  \\
\hline
\end{tabular*}
\end{center}

The boldfaced results in this table show that commuting a non-diagonal matrix $\lambda_i$ with a diagonal matrix ($\lambda_3$ or $\lambda_8$) gives a result proportional to $\lambda_i$. This property is very useful while quantizing $SU(3)$ gauge fields. Mathematically, we can put it into the form
\begin{equation}
{f^{ij}}_k = 0 \quad \quad \text{if} \quad \quad i=3,8 \quad \text{and} \quad j,k=1,2,4,5,6,7, \quad j\neq k.
\end{equation}

\section{Matter current in $SU(3)$}

\hspace{0.5cm} We want to show some of the results we arrive at, when we expand components of the matter current in terms of creation and annihilation operators.

For diagonal components, we have
\begin{eqnarray*}
J_3         & = & \sum_{m,n=\frac{1}{2}}^\infty \left\{ \delta_{m+k}^n \left[
                   b_{m,2}^\dagger b_{n,2}
                       \left(\frac{u_{m,2}}{u_{n,2}}+\frac{u_{n,2}}{u_{m,2}} \right)
                  -\frac{1}{2} b_{m,4}^\dagger b_{n,4}
                       \left(\frac{u_{m,4}}{u_{n,4}}+\frac{u_{n,4}}{u_{m,4}} \right)
                  \right. \right. \\
            &   & \quad \quad\quad \quad\quad\quad \quad\quad \quad\quad \quad\quad \quad
                  \quad \quad\quad \quad \left.
                  -\frac{1}{2} b_{m,7}^\dagger b_{n,7}
                       \left(\frac{u_{m,7}}{u_{n,7}}+\frac{u_{n,7}}{u_{m,7}} \right)
                  \right] \\
            &   & \quad \quad \quad- \delta_{m+k}^{-n} \left[
                  b_{m,2}^\dagger d_{n,2}^\dagger
                       \left(\frac{u_{m,2}}{v_{n,2}}-\frac{v_{n,2}}{u_{m,2}} \right)
                  -\frac{1}{2} b_{m,4}^\dagger d_{n,4}^\dagger
                       \left(\frac{u_{m,4}}{v_{n,4}}-\frac{v_{n,4}}{u_{m,4}} \right)
                  \right. \\
            &   & \quad \quad\quad\quad \quad\quad \quad\quad \quad\quad \quad\quad \quad
                  \quad \quad\quad \quad \left.
                  -\frac{1}{2} b_{m,7}^\dagger d_{n,7}^\dagger
                       \left(\frac{u_{m,7}}{v_{n,7}}-\frac{v_{n,7}}{u_{m,7}} \right)
                  \right] \\
            &   & \quad \quad\quad - \delta_{m+k}^n \left[
                  d_{m,2}^\dagger d_{n,2}
                       \left(\frac{v_{m,2}}{v_{n,2}}+\frac{v_{n,2}}{v_{m,2}} \right)
                  -\frac{1}{2} d_{m,4}^\dagger d_{n,4}
                       \left(\frac{v_{m,4}}{v_{n,4}}+\frac{v_{n,4}}{v_{m,4}} \right)
                 \right. \\
            &   & \quad\quad \quad\quad \quad\quad \quad\quad \quad\quad \quad\quad \quad
                  \quad \quad\quad \quad \left.
                  -\frac{1}{2} d_{m,7}^\dagger d_{n,7}
                       \left(\frac{v_{m,7}}{v_{n,7}}+\frac{v_{n,7}}{v_{m,7}} \right)
                  \right] \\
            &   & \quad\quad \quad - \delta_{m+n}^k \left[
                  b_{m,2} d_{n,2}
                    \left(\frac{u_{m,2}}{v_{n,2}}-\frac{v_{n,2}}{u_{m,2}} \right)
                  -\frac{1}{2} b_{m,4} d_{n,4}
                    \left(\frac{u_{m,4}}{v_{n,4}}-\frac{v_{n,4}}{u_{m,4}} \right)
                 \right. \\
            &   & \quad \quad\quad \quad\quad \quad\quad \quad\quad \quad\quad \quad \quad
                  \quad \quad\quad \quad \left. \left.
                  -\frac{1}{2} b_{m,7} d_{n,7}
                    \left(\frac{u_{m,7}}{v_{n,7}}-\frac{v_{n,7}}{u_{m,7}} \right)
                  \right] \right\}
\end{eqnarray*}
and
\begin{eqnarray*}
J_8         & = & \sum_{m,n=\frac{1}{2}}^\infty \left\{ \delta_{m+k}^n \left[
                   b_{m,7}^\dagger b_{n,7}
                       \left(\frac{u_{m,7}}{u_{n,7}}+\frac{u_{n,7}}{u_{m,7}} \right)
                  - b_{m,4}^\dagger b_{n,4}
                       \left(\frac{u_{m,4}}{u_{n,4}}+\frac{u_{n,4}}{u_{m,4}} \right)
                  \right] \right. \\
            &   & \quad \quad \quad- \delta_{m+k}^{-n} \left[
                  b_{m,7}^\dagger d_{n,7}^\dagger
                       \left(\frac{u_{m,7}}{v_{n,7}}-\frac{v_{n,7}}{u_{m,7}} \right)
                  - b_{m,4}^\dagger d_{n,4}^\dagger
                       \left(\frac{u_{m,4}}{v_{n,4}}-\frac{v_{n,4}}{u_{m,4}} \right)
                  \right] \\
            &   & \quad \quad\quad - \delta_{m+k}^n \left[
                  d_{m,7}^\dagger d_{n,7}
                       \left(\frac{v_{m,7}}{v_{n,7}}+\frac{v_{n,7}}{v_{m,7}} \right)
                  -d_{m,4}^\dagger d_{n,4}
                       \left(\frac{v_{m,4}}{v_{n,4}}+\frac{v_{n,4}}{v_{m,4}} \right)
                  \right] \\
            &   & \quad\quad \quad - \delta_{m+n}^k \left.\left[
                  b_{m,7} d_{n,7}
                    \left(\frac{u_{m,7}}{v_{n,7}}-\frac{v_{n,7}}{u_{m,7}} \right)
                  - b_{m,4} d_{n,4}
                    \left(\frac{u_{m,4}}{v_{n,4}}-\frac{v_{n,4}}{u_{m,4}} \right)
                  \right] \right\}.
\end{eqnarray*}

The off-diagonal components may be written as the sum of a term depending only on normal modes, and a term with the zero-modes. For example, for $J_1$, the first part is found to be

\begin{eqnarray*}
Q_1(k) & = & \sum_{m=\frac{1}{2}}^\infty \sum_{n=1}^\infty
              \left( \frac{u_{m,2}}{w_{m,3}} -\frac{w_{m,3}}{u_{m,2}} \right)
              a_{n,3}^\dagger b_{m,2}^\dagger \delta_{m+n}^{-k}
              -\left( \frac{v_{m,2}}{w_{m,3}} -\frac{w_{m,3}}{v_{m,2}} \right)
              a_{n,3} d_{m,2} \delta_{m+n}^k \\
       &   & \quad \quad \quad
              -\left( \frac{u_{m,2}}{w_{m,3}} +\frac{w_{m,3}}{u_{m,2}} \right)
              a_{n,3} b_{m,2}^\dagger \delta_{m+k}^n
              +\left( \frac{v_{m,2}}{w_{m,3}} -\frac{w_{m,3}}{v_{m,2}} \right)
              a_{n,3}^\dagger d_{m,2} \delta_{k+n}^m \\
       & + & \frac{1}{\sqrt{2}} \sum_{m=\frac{1}{2}}^\infty \sum_{n=1}^\infty
             \left( \frac{u_{n,4}}{u_{m,7}} - \frac{u_{m,7}}{u_{n,4}} \right)
             b_{m,7} b_{n,4} \delta_{m+n-m_{0,4}-m_{0,7}}^{k+m_{0,2}} \\
       &   & \quad \quad \quad\quad
           + \left( \frac{v_{n,4}}{u_{m,7}} + \frac{u_{m,7}}{v_{n,4}} \right)
             b_{m,7} d_{n,4}^\dagger \delta_{m-n-m_{0,4}-m_{0,7}}^{k+m_{0,2}} \\
       &   & \quad \quad \quad\quad
           - \left( \frac{u_{n,4}}{v_{m,7}} + \frac{v_{m,7}}{u_{n,4}} \right)
             d_{m,7}^\dagger d_{n,4} \delta_{n-m-m_{0,4}-m_{0,7}}^{k+m_{0,2}} \\
       &   & \quad \quad \quad\quad
           - \left( \frac{v_{n,4}}{v_{m,7}} - \frac{v_{m,7}}{v_{n,4}} \right)
             d_{m,7}^\dagger d_{n,4}^\dagger \delta_{-m-n-m_{0,4}-m_{0,7}}^{k+m_{0,2}}.
\end{eqnarray*}
Is is interesting to see that the four last terms in this sum are not present in the $SU(2)$ case. They are independent of the $a$-type particles.

\end{appendix}

\end{document}